\documentclass[twocolumn,showpacs,preprintnumbers,amsmath,amssymb]{revtex4-1}
\usepackage{graphicx}
\usepackage{dcolumn}
\usepackage{bm}
\newcommand {\be}{\begin{equation}}
\newcommand {\ee}{\end{equation}}
\newcommand {\bea}{\begin{eqnarray}}
\newcommand {\eea}{\end{eqnarray}}
\newcommand {\bem}{\begin{displaymath}}
\newcommand {\eem}{\end{displaymath}}
\newcommand {\p}{\partial}
\newcommand {\f}{\frac }

\begin{document}
\title{Energy Transport in Closed Quantum Systems}
\author{ G. A. Levin, W. A. Jones, K. Walczak, and K. L. Yerkes \\}
\affiliation{ Propulsion Directorate, Air Force Research Laboratory, Wright-Patterson Air Force Base, OH 45433}
\date{\today}
\begin{abstract}
We examine energy transport in an ensemble of closed quantum systems driven by stochastic perturbations. 
One can show that the probability and energy fluxes can be described in terms of quantum advection modes (QAM) 
associated with the off-diagonal elements of the density matrix. These QAM play the role of Landauer channels in a system with discrete energy spectrum and the eigenfunctions that cannot be described as plane waves. In order to determine the type of correlations that exist between the direction and magnitudes of each QAM and the average direction of energy and probability fluxes we have numerically solved the time-dependent Schr\"{o}dinger equation describing a single 
particle trapped in a parabolic potential well which is perturbed by stochastic 'ripples'. 
The ripples serve as a localized energy source and are offset to one side of the potential well. 
As the result a non-zero net energy flux flows from one part of the potential well to another across the symmetry center of the
potential. We find that some modes exhibit positive correlation with the direction of the energy flow. Other modes, that carry a smaller energy per unit of the probability flux, anticorrelate with the energy flow
and thus provide a backflow of the probability. The overall picture of energy transport that emerges from our results is very different from the conventional one based on a system with continuous energy spectrum. 
\end{abstract}
\maketitle
\section{\label{sec:level1}Introduction\protect}

Understanding the properties of the energy transport by quantum channels is one of the most important problems of non-equilibrium thermodynamics, important from the fundamental point of view as well as for its applications. 
A typical problem addressed in literature is the energy transport between two thermal reservoirs coupled by a quantum system (a mediator).
The description of energy flux is usually stated in terms of Landauer channels\cite{Lan1,Anderson,Imry,Lepri} described as plane waves 
that carry energy and probability flux in a given direction with occupation numbers determined by the temperature of the reservoir from which a given channel originates. 

In this paper we address the question of energy flux from a different perspective. One can show that both probability and energy fluxes can be described as superpositions of quantum advection modes (QAM). 
These modes are associated with the off-diagonal elements of the density matrix appropriately defined for our ensemble.
Each such a mode carries a certain probability flux and a fixed, quantized amount of energy per unit of the probability flux.
In our opinion, these modes are the prototypes of what is commonly known as Landauer transport channels. 

In order to determine the statistical properties of QAM and how they contribute to the overall energy and probability flux we
have considered a model of a particle confined within a potential well and described by the Schr\"{o}dinger equation. We introduce a stochastic time-dependent perturbation of the confining potential ("ripples") that has certain well defined statistical properties. There is no feedback from the confined particle to the source of the ripples, so that the net result of the perturbation is a steady increase in energy of the confined particle. 

Since the power source is localized, there is a nonzero average energy flow from one spatial region of the confining potential to another.
Thus, we can study the statistical properties of the energy and probability fluxes without addressing a more involved problem of two 
thermal reservoirs entangled with the mediator. 
This model is somewhat similar to that describing trapped ions and atoms where electromagnetic perturbations lead to gradual heating of the confined particles \cite{Turchette, Turchette2, Leib}. 

From the solution $\Psi (\vec{r},t)$ of the Schr\"{o}dinger equation, which includes perturbation by ripples, one can calculate the probability flux $\vec{j}_p(\vec{r},t)$. Similarly, one can introduce the energy density $\epsilon (\vec{r},t)$ and the energy flux $\vec{j}_E(\vec{r},t)$, both expressed in terms of $\Psi (\vec{r},t)$ and its time derivative $\dot{\Psi} (\vec{r},t)$, that satisfy the energy conservation condition in the form of the continuity equation. This equation includes also the energy source determined by the time-dependent perturbation. 
The source of stochasticity in our approach is a {\it classical} Brownian particle immersed in a thermal bath
which is characterized by temperature T. The motion of the Brownian particle creates ripples in the confining potential.
The spatial trajectory of the Brownian particle is determined by the standard Langevin equation. Thus, the quantum particle described by the Schr\"{o}dinger equation is affected by the movement of the classical Brownian particle without any feedback. 

Our approach is as follows: We generate, solving numerically the Langevin equation, a time dependent trajectory of the classical Brownian particle. Then, feeding this trajectory into the Schr\"{o}dinger equation as part of the perturbation potential, we obtain the solution $\Psi (\vec{r},t)$. After that we start the procedure all over again, generating another random trajectory and another solution $\Psi (\vec{r},t)$. By repeating this process many times we accumulate an ensemble consisting of the wave functions
$\Psi (\vec{r},t)$. Each of these solutions has evolved under the action of the perturbations, all of which have the same statistical properties determined by the movement of the Brownian particle. 
It is important to emphasize that this ensemble is not a traditional thermodynamic ensemble, 
it is rather a collection of evolving pure quantum states. 

For each member of this ensemble we calculate probability and energy fluxes and analyze the statistical properties of 
the resulting ensemble of fluxes. 

This paper is structured as follows. Section II defines the quantum advection modes in terms of the off-diagonal elements of the density matrix.
Section III introduces the model which we use to determine the statistical properties of these modes. Section IV presents the results which include the distributions of the probability and energy fluxes. 
The ensemble average energy flux is treated as a superposition of two statistically orthogonal fluxes. One is a purely advective energy flux fully correlated with the probability flux. The other component - the thermal flux - has nonzero average value and corresponds to the energy flow uncorrelated with the movement of the particle.   
The subsection IV(B) presents the correlations between the direction of the individual QAM and the direction of the energy flow averaged over the time intervals where the net probability flux is zero.
\section{\label{sec:level1}Quantum Advection Modes\protect}
Let us first consider some general statements regarding energy flux that corresponds to a wave function $\Psi (\vec{r},t)$
satisfying a time-dependent Schr\"{o}dinger equation:
\be
i\hbar\f{\p\Psi}{\p t}= \hat{H}\Psi;\;
\hat{H}=\f{\hat{p}^2}{2m}+V(\vec{r},t),
\ee
where $\hat {p}\equiv -i\hbar\nabla $ and the potential $V(\vec{r},t)=V_0(\vec{r})+U(\vec{r},t)$ consists of a confining potential $V_0(\vec{r})$ and a superimposed upon it time dependent perturbation $U$. 
The probability of finding a particle in a given place $P(\vec{r},t)=|\Psi (\vec{r},t)|^2$ obeys the continuity equation
\be
\f{\p P }{\p t}+\nabla\cdot \vec{j}_p=0,
\ee
where the probability flux $\vec{j}_p$ is given by
\be
\vec{j}_p=\f{\hbar}{m}\Im (\Psi^{\ast }\nabla\Psi ).
\ee

The energy density can be defined as follows
\be
\epsilon (\vec{r},t) =\f{1}{2}[\Psi^{\ast}(\hat{H}\Psi )+(\hat{H}\Psi^{\ast})\Psi ]=
\Re (\Psi^{\ast}\hat{H}\Psi ).
\ee
It is a real number, which after integration over the total volume gives the expectation value of the Hamiltonian 
\be
E_{tot}\equiv \int \epsilon (\vec{r},t) d\vec{r}=\langle\hat{H}\rangle .
\ee
By virtue of Eq. (1) the energy density can be written in the form
\be
\epsilon (\vec{r},t) = -\hbar \Im (\Psi^{\ast}\dot{\Psi }).
\ee
The quantity defined by Eq. (6) was introduced by D. Bohm \cite{Bohm}. He suggested to readers to prove that it satisfies the conservation law (in the case of time-independent Hamiltonian). At the risk of upsetting future students by denying them a pleasure to derive the energy continuity equation themselves, we give in the Appendix a brief derivation of the following equation
\be
\f{\p\epsilon }{\p t}+\nabla\cdot\vec {j}_E =Q(\vec{r},t).
\ee
Here $\vec{j}_E$ is the energy flux 
\be
\vec{j}_E= \f{\hbar}{2m}\Im \left\{\Psi^{\ast }\nabla (\hat{H}\Psi )-
(\hat{H}\Psi ) \nabla \Psi^{\ast}\right\}.
\ee
A more convenient for numerical calculations form is
\be
\vec{j}_E= \f{\hbar^2}{2m}\Re \left\{\Psi^{\ast }\nabla\dot{\Psi }-
\dot{\Psi } \nabla \Psi^{\ast}\right\}.
\ee
The energy source/sink in Eq. (7) is determined by the time dependent part of the Hamiltonian,
\be
Q=\dot{V}|\Psi |^2.
\ee

Any solution of Eq. (1) can be presented in a form
\be
\Psi(\vec{r},t)=\sum_n c_n(t)e^{-i\omega_n t}\psi_n(\vec{r});\;\;\omega_n\equiv E_n/\hbar.
\ee
Here we consider the wavefunctions $\psi_n(\vec{r})$ to be the eigenstates of the unperturbed Hamiltonian 
\be
\hat{H}_0=\f{\hat{p}^2}{2m}+V_0(\vec{r});\;\; \hat{H}_0\psi_n=E_n\psi_n.
\ee
Substituting (11) in Eq. (3) we obtain the probability flux expressed in terms of the stochastic coefficients $c_n(t)$
\be 
\vec{j}_p (\vec{r},t)=\f{\hbar}{m}\sum_{n>m} \Im \left(c_nc_m^{\ast} e^{-i\omega_{nm} t}\right )\vec{g}_{nm}(\vec{r}),
\ee
where $\omega_{nm}\equiv \omega_n-\omega_m \ge 0$. The spatial distribution of the flux is determined by the generalized Wronskians of the two wave functions 
\be
\vec{g}_{nm}(\vec{r}) =\nabla\psi_n(\vec{r})\psi_m(\vec{r}) - \psi_n(\vec{r})\nabla\psi_m(\vec{r}).
\ee

This expression for the probability flux is universal even though the wave functions $\psi_n(\vec{r})$ are not the eigenstates of the full Hamiltonian $\hat{H}$. It is a bit more complicated for the energy flux. Let us consider a situation when the time dependent perturbation $U(\vec{r},t)$ is localized, so that there are spatial regions within the confining potential free of perturbation. In these regions
\be
\hat{H}(\vec{r},t)\approx \hat{H}_0(\vec{r})
\ee
and, correspondingly,
\be
\hat{H}(\vec{r},t)\psi_n(\vec{r})\approx E_n\psi_n(\vec{r}).
\ee
Within these free of perturbation regions we obtain from Eqs. (8) and (11)
\be
\vec{j}_E (\vec{r},t)=\f{\hbar }{m}\sum_{n>m}\f{E_n+E_m}{2} \Im \left(c_nc_m^{\ast} e^{-i\omega_{nm} t}\right )\vec{g}_{nm}(\vec{r}).
\ee
Thus,
\bea
\vec{j}_p (\vec{r},t)=\sum_{n>m}\vec{q}_{nm} (\vec{r},t);\; \\ \nonumber
\vec{j}_E (\vec{r},t)=\sum_{n>m}\epsilon_{nm}\vec{q}_{nm} (\vec{r},t).
\eea
Each of the quantum advection modes (QAM) 
\be
\vec{q}_{nm} (\vec{r},t)=\f{\hbar}{m} \Im \left(c_nc_m^{\ast} e^{-i\omega_{nm} t}\right )\vec{g}_{nm}(\vec{r})
\ee
carries a certain amount of the probability flux determined by the correlator 
$(c_nc_m^{\ast} e^{-i\omega_{nm} t})$. We will loosely call these correlators off-diagonal elements of the density matrix, even though 
the ensemble over which the averaging takes place is not necessarily a conventional thermodynamic ensemble. The amount of energy carried by each mode per unit of the probability flux is fixed, given by the average energy of the two levels involved,
\be
\epsilon_{nm}=\f{E_n+E_m}{2}.
\ee

Hereafter we will ignore the spin degeneracy. We will call two modes $\vec{q}_{nm}$ and $ \vec{q}_{n^{\prime}m^{\prime }}$ degenerate if they carry the same energy per unit of the probability flux, namely if $\epsilon_{nm}=\epsilon_{n^{\prime}m^{\prime}}$.
Consider, as an example, one-dimensional oscillator with equidistant energy levels
$E_n=\hbar\omega (n+1/2)$, so that $\epsilon_{nm}=\hbar\omega (n+m+1)/2$. 
The lowest energy mode with positive spatial parity $q_{10}$ is nondegenerate and carries one quantum of energy per unit of the probability flux. The lowest energy mode with negative parity $q_{20}$ carries fractional amount $3\hbar\omega /2$
per unit of the probability flux and is also nondegenerate. Two degenerate modes, both of positive parity, $q_{30}$ and $q_{21}$ carry two quanta of energy per unit of the probability flux, etc. 

A few immediately obvious conclusions follow from this analysis. 
Each QAM corresponds to purely advective energy transfer, namely the energy flux carried by each mode is directly proportional to the probability flux. However, in a system where the net average probability flux is zero (there is no net mass or charge transfer) it is possible to have a non-zero energy flow, provided that at least two non-degenerate QAM are activated and both have nonzero average, so that $\vec{j}_p=\vec{q}_{\alpha} +\vec{q}_{\beta }=0$, but 
$\vec{j}_E=\epsilon_{\alpha}\vec{q}_{\alpha} +\epsilon_{\beta}\vec{q}_{\beta }\neq 0$. 
Two nondegenerate modes require at least three energy levels involved. A two-level system cannot conduct energy without carrying nonzero probability (mass or charge ) flux. The energy flow unaccompanied by mass flow is usually defined as thermal conduction (heat flux). 

It is important to realize also that energy flow through a quantum system (such as a qubit) establishes a degree of persistent coherence between several energy levels, so that at least some correlators in (19) do not average to zero for as long as the external conditions that create the energy flow are maintained. 

\section{\label{sec:level1} Model\protect}

In order to determine the properties of the quantum advection modes we have studied a following model. 
We have solve numerically the one-dimensional Schr\"{o}dinger equation (1) with parabolic confinement potential
\be
V_0(x)=\f{m\omega^2x^2}{2}.
\ee
A time dependent perturbation is introduced as a fixed shape "protrusion", $U(x- x_1 (t))$, riding on the back of a classical Brownian particle whose trajectory is $x_1 (t)$, Fig. 1(a). Specifically, we choose
\be
U(x- x_1 (t))=U_0\exp \left \{ -\f{(x-x_1)^2}{2\delta^2} \right \}.
\ee
The classical Brownian particle is trapped in its own harmonic potential well and is in equilibrium with the thermal bath of temperature $T$. It is important to emphasize that the center of the potential well in which the Brownian particle is trapped is located off center of the potential well $V_0(x)$, as shown in Fig. 1(a). We can choose parameters, such as the location of the center of the trap and the stiffness constant of the trap in such a way that most of the time the Brownian particle and, correspondingly, the perturbation are localized on the right-hand side, $x>0$, of the confining potential $V_0(x)$. Thus, the left-hand side of $V_0(x)$ is mostly free of perturbation, which is a condition, Eqs. (15) and (16), necessary for defining the energy flux in terms of the quantum advection modes, Eqs. (17) and (18). 

To make the perturbation a ripple, rather than a protrusion, we define it by subtracting the static part $U(x-a)$. 
Thus, the potential in Eq. (1) is defined as
\bea
V(x,t)=V_0(x)+ \delta U(x,t);\;\\ \nonumber
\delta U(x,t) \equiv U(x-x_1(t))-U(x-a).
\eea
Here $a>0$ is the location of the center of the trap for the Brownian particle. If the Brownian particle were at rest at the bottom of its trap, 
$\delta U\equiv 0$. Since its motion is determined by the temperature of the thermal reservoir, the perturbation vanishes at $T=0$.
At all times $\int_{-\infty }^{\infty } \delta U(x,t)dx \equiv 0$. Figure 1(b) shows a snapshot of the total potential $V_0(x)+\delta U(x,t)$.

Thus, we solve numerically the folowing Schr\"{o}dinger equation
\be
i\hbar\f{\p\Psi}{\p t}=-\f{\hbar^2}{2m}\f{\p^2\Psi}{\p x^2}+
(V_0(x) +\delta U(x,t))\Psi.
\ee
\begin{figure}
\includegraphics{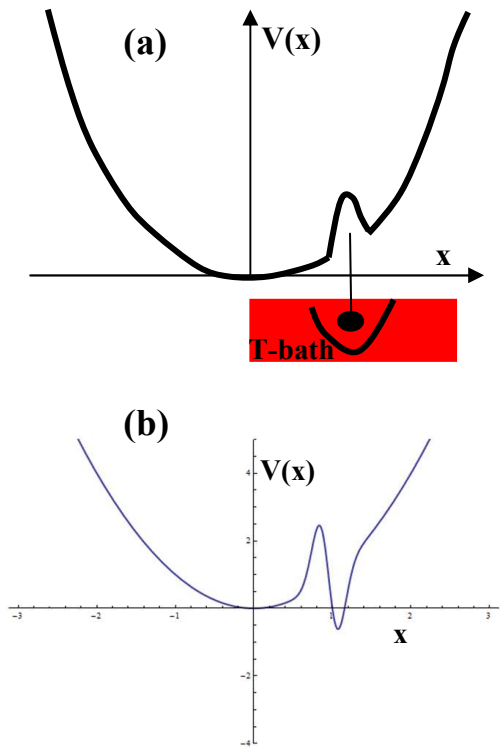}
\caption{\label{fig:} (\bf{a}) A sketch of the potential $V(x,t)$ consisting of a static, symmetric confining potential with superimposed protrusion driven by a classical Brownian particle immersed in a thermal bath. (\bf{b}) A snapshot of the potential $V(x,t)$, Eq. (23).}
\end{figure}

A trajectory of the Brownian particle necessary to define the perturbation (23) is obtained by solving numerically the Langevin equation which determines the stochastic trajectories $x_1(t)$, where 
\be
M\ddot {x}_1+\eta \dot{x}_1+M\omega_1^2(x_1-a)=F,
\ee
with the random force correlator
\be
\langle F(t)F(t^{\prime})\rangle=2\eta k_BT\delta (t-t^{\prime}).
\ee

Our approach to analyzing the processes caused by such stochastic perturbation is as follows. A trajectory $x_1(t)$ is generated by solving numerically Eqs. (25, 26). This trajectory is then fed into the numerical algorithm which solves Eq. (24) with the potential given by (22, 23). 
This gives us the wave function $\Psi_1 (x, t)$. Then another trajectory $x_1(t)$ is generated by the same equations (25, 26) and the second solution of Eq. (24) $\Psi_2(x,t)$ is obtained. This protocol is repeated $N$ times. As the result we obtain an ensemble of solutions $\{\Psi_1(x,t) \ldots\Psi_N(x,t)\}$ which allows us to determine the ensemble- and time-average, as well as the distributions, of different physical quantities such as the probability and energy fluxes, the density matrix elements, etc. The data presented here are the result of averaging over $N=10^4$ copies of the quantum system exposed to the action of the stochastic ripples. It should be noted that this is an ensemble of pure quantum states, so that it contains all information about the quantum system, rather than a truncated amount of information described by the conventional density matrix. The total energy of this ensemble is not fixed, in fact it increases at a certain rate. In order to analyze our results we do not introduce any "average" wave function. We are only averaging the probabilities and fluxes, i.e. the quadratic forms of the individual wave functions that constitute this ensemble. 

\subsection{\label{sec:level1}Implementation\protect}

In the appropriate dimensionless variables
\be
\tau=\omega t; \;\xi=x/\ell,
\ee
Eq. (24) takes form
\be
i\f{\p\Psi}{\p \tau}=-\f{1}{2}\f{\p^2\Psi}{\p \xi^2}+
\left (\f{\xi^2}{2} +u(\xi ,\tau)\right )\Psi,
\ee
where 
\be
\ell=\left (\f{\hbar}{m\omega} \right )^{1/2}; \; (m\omega^2\ell^2=\hbar \omega).
\ee 
Energies and temperature are measured in their natural unit $\hbar\omega $, so that
\be
u(\xi,\tau )=\f{\delta U(\xi,\tau )}{\hbar\omega}.
\ee
The unperturbed energy levels are given by
\be
E_n=n+1/2,
\ee
and dimensionless temperature is defined as
\be
\theta = \f{k_BT}{\hbar \omega}.
\ee
The Langevin equation (25) takes form
\be
\ddot {\xi_1}+\gamma \dot{\xi_1}+\Omega^2\xi_1=f(\tau ),
\ee
where $\xi_1=(x_1-a)/\ell$. The relative frequency of the confining potential and the relative damping rate
are given by
\be
\Omega^2\equiv \f{\omega_1^2}{\omega^2};\; \gamma\equiv \f{\eta}{M\omega}.
\ee
The third parameter describing the Brownian particle is the relative stiffness of its trap
\be
\kappa \equiv \f{k}{k_1}\equiv \f{m\omega^2}{M\omega_1^2}
\ee
which determines the characteristic scale of the displacement of the Brownian particle from the center of its trap
\be
\ell_1^2=\f{k_BT}{k_1}; \;\; \f{\ell_1^2}{\ell^2}=\kappa\theta.
\ee
The random force in Eq. (33) is defined as
\be
f=\f{F}{M\ell\omega^2}
\ee
with the correlator
\be
\langle f(\tau)f(\tau^{\prime})\rangle=2\gamma\theta(\kappa\Omega^2)\delta (\tau-\tau^{\prime}).
\ee

The Langevin equation (25, 26) describes a particle in thermal equilibrium. Therefore, the trajectories obtained as its solutions 
must have the statistical properties defined by the Boltzmann probability distributions for the displacement and velocity 
\be
P(x_1)=\left (\f{M\omega_1^2}{2\pi k_BT}\right )^{1/2}
\exp\left \{-\f{M\omega_1^2(x_1-a)^2}{2k_BT}\right \},
\ee
and
\be
P(\dot{x}_1)=\left (\f{M}{2\pi k_BT}\right )^{1/2}
\exp\left \{-\f{M\dot{x_1}^2}{2k_BT}\right \}.
\ee
Here $P(x_1)dx_1$ and $P(\dot{x}_1)d\dot{x}_1$ are relative frequencies with which the trajectories acquire the 
respective values. 
We have used these conditions as a way to validate the correctness of our numerical solutions of the Langevin equation. 

The numerical parameters that were used in this study were as follows. The temperature range was $\theta \sim 1$. The frequency of the trap $\omega_1$ was chosen close to the main frequency $\omega$, and the damping rate $\gamma$ was of the same order of magnitude as 
$\Omega$. The data set shown below correspond to
\be
\Omega = 1.1;\; \gamma =1.6.
\ee
If we choose the range of parameters substantially different from that shown above, the perturbation will be either too weak or too strong to obtain reliable results using numerical methods of solution of the Schr\"{o}dinger equation. 

We also need to ensure that the probability of the perturbation (23) to extend to the left-hand side ($x\le 0$) of the confining potential $V_0(x)$ is statistically negligible. This is necessary to meet conditions (15, 16) so that we can use the formalism of QAM to describe the transport in the left-hand side of the potential well. For that purpose the parameters of the trap displacement $a$ and the stiffness of the trap have to be chosen so that the probability to find the Brownian particle at $x_1<0$ was negligibly small. The probability (39) to find the Brownian particle at $x_1=0$ is given by
\be
\left. P\right |_{x_1=0}=
\left (\f{1}{2\pi \kappa\theta}\right )^{1/2}
\exp\{-a^2/2\ell^2\kappa\theta\}.
\ee
In our numerical calculations we have used the values 
\be
a/\ell=1;\; \kappa=1/36.
\ee
If in Eq. (23) the displacement $a/\ell$ is substantially greater than $1$, it reduces exponentially the coupling constants between the energy levels. The stiffness of the trap $36$ times greater than that of the potential $V_0$, coupled with the width of the Gaussian (22)
\be
\delta = 1/8
\ee
ensures that the left-hand side of the potential well $V_0(x)$ is practically free of perturbation.
\subsection{\label{sec:level1}Numerical methods\protect}

To solve the Langevin equation given by Eq. (33), we used the standard fourth-order Runge-Kutta (RK4) iterative method. 
The random force, Eq. (38), was defined as follows:
\be
f(t_i)=\f{\eta (t_i)}{\Delta t^{1/2}},
\ee
where $\Delta t$ is the time step and $\eta$ is a pseudo-random variable distributed with the probability
\be
P(\eta )=\left (\f{1}{4\pi\gamma\theta \kappa\Omega^2 }\right )^{1/2} \exp\{-\eta^2/4\gamma\theta \kappa\Omega^2\}.
\ee
Validation of the algorithm was carried out by comparing the statistical properties of the trajectories with conditions (39, 40). 

The solution for time-dependent Schr\"{o}dinger equation given by Eq.(28)
is obtained using the Crank-Nicolson method, similar to that discussed in \cite{Baskakov,Fijany,Sarwari}.
In each individual run the perturbation was turned on and turned off gradually at the beginning and the end of the run in order to avoid 
the excitation of too many energy levels.
Validation of the algorithm was carried out by controlling the norm of the wave function and by running the 
algorithm without perturbation and comparing the results with the exact solution. It was determined that error 
greater than $5\%$ accumulates after approximately $20-30$ classical periods $2\pi/\omega$. 
This places the limit on the duration of time that we have used to run numerical calculation for the individual solution 
$\Psi (x, t)$. 
\section{\label{sec:level1}Results\protect}
\begin{figure}
\begin{tabular}{cc}
\includegraphics [width=.9\columnwidth]{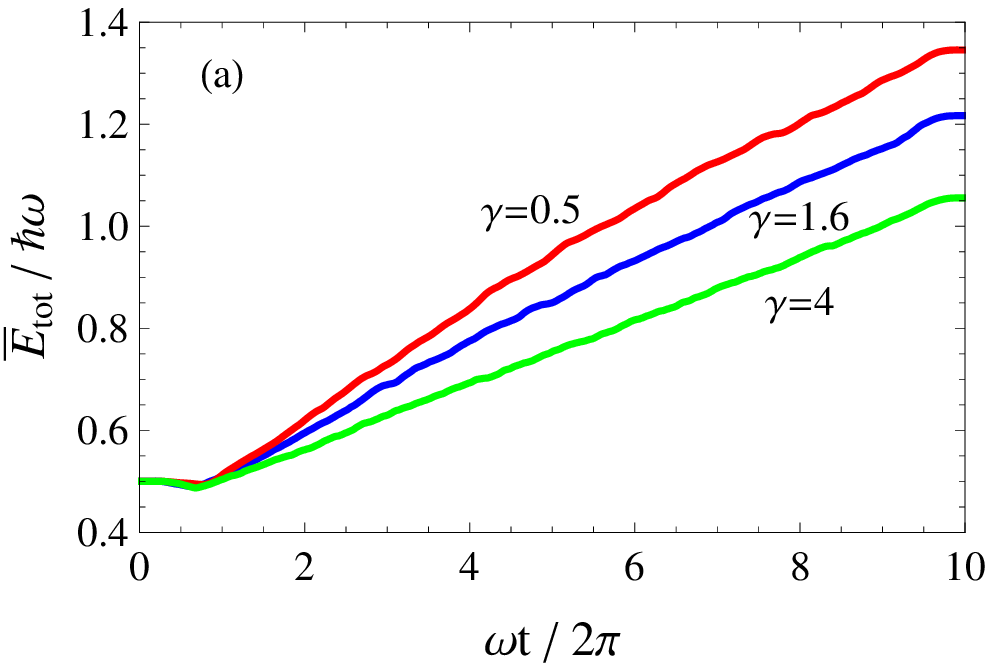} \\
\includegraphics [width=.9\columnwidth]{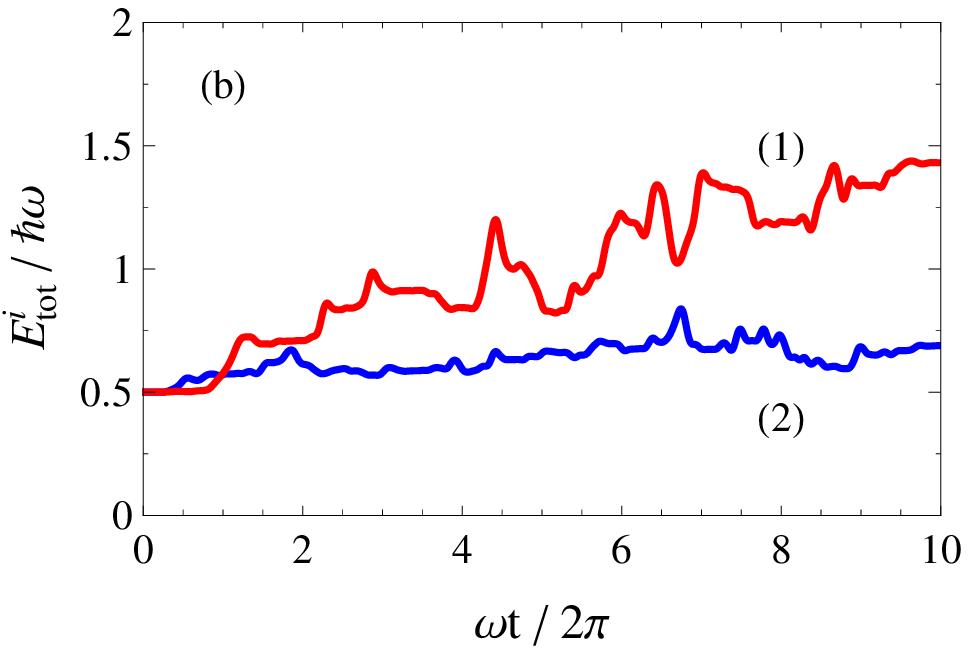}
\end{tabular}
\caption{\label{fig:} (\bf{a}) Ensemble average energy, Eq. (47), for three different values of the friction coefficient $\gamma$ in Eq. (33). (\bf{b}) Two examples of the energy of an individual member of the ensemble $E_{tot}^i$ . In contrast to the ensemble average energy, the energy of an individual quantum system can decrease over some time intervals.}
\end{figure}

Figure 2 (a) shows the ensemble average energy 
\be
\bar{E}_{tot}=N^{-1}\sum_{i=1}^{N}E_{tot}^i
\ee
obtained by averaging over $N=10^4$ solutions $\Psi_i(x,t)$ of Eq. (24). The individual energies $ E_{tot}^i$ were calculated according to Eqs. (5, 6)
\be
E_{tot}^i(t)=-\hbar \int_{-\infty}^{\infty}\Im (\Psi_i^{\ast}(x,t)\dot{\Psi}_i (x,t))dx.
\ee

All results shown in this paper correspond to the ground state initial condition for  all members of the ensemble, so that  $E_{tot}^i(0)/\hbar\omega=1/2$.
On average, the perturbation continuously pumps energy into the system. This is true in the case of a spatially uniform perturbations (heating of trapped ions) \cite{Savard,Turchette} and in our model as well. The rate of energy increase depends on the parameters of the Brownian particle. Other things being equal, the lower friction leads to greater rate of energy increase. 
However, the averaging conceals the full picture of this phenomenon. Figure 2 (b) shows the energy of two arbitrary chosen individual members of the ensemble.
On the level of an individual quantum system the energy does not monotonically increase. There are intervals of time when the perturbation 
serves as the energy drain. 

\subsection{\label{sec:level1}Probability and Energy Fluxes\protect}
\begin{figure}
\includegraphics[width=.9\columnwidth]{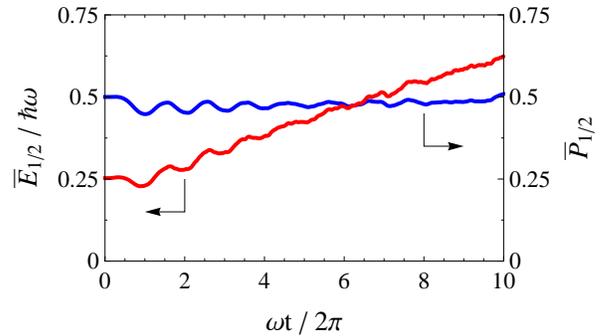}
\caption{\label{fig:} (\bf{a}) Time dependence of the ensemble average $\bar{E}_{1/2}$ and $\bar{P}_{1/2}$.}
\end{figure}

In our model there are no two heat reservoirs between which energy can flow.
Nevertheless, it is easy to see that this model does allow a persistent energy flow in the absense of the net probability flux. 
Let us consider a "half-bucket" energy
\be
E_{1/2}\equiv \int_{-\infty}^0\epsilon (x,t)dx.
\ee
If we integrate Eq. (7), we get 
\be
\left. \f{dE_{1/2}}{dt}=- j_E (t)\right |_{x=0}.
\ee
Here we have taken into account that the energy source $Q(x,t)$ is localized at $x>0$, so that the energy inflow into the $x\leq 0$ half of the confinement is due to the energy flux across the centerline. In other words, the energy flux can be calculated on the basis of Eq. (9) or, in the spatial regions free of perturbation, on the basis of Eq. (50). 

Figure 3 shows the ensemble average $\bar{E}_{1/2}$. The averaging procedure is the same as in (47). 
Obviously, the excess of energy introduced by the perturbation does not accumulate in one half of the confining potential, so that $\bar{E}_{1/2}\approx \bar{E}_{tot}/2$ almost monotonically increases and, therefore, there is a persistent net energy flux across the centerline (from right to left) $\bar{j}_E <0$ . 
\begin{figure*}
\includegraphics{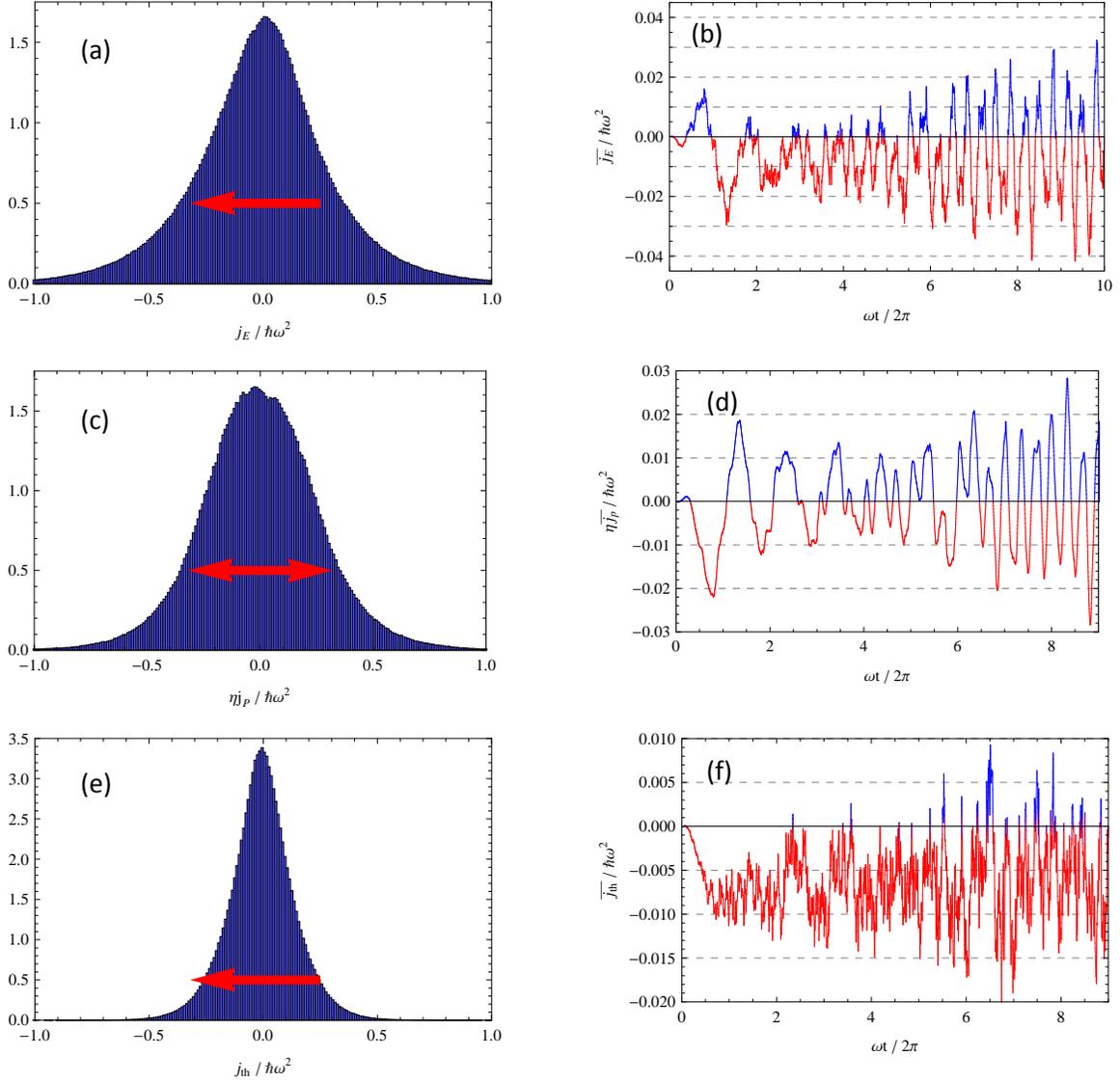}
\caption{\label{fig:} (\bf{a}) The histogram shows the distribution of the values of the energy flux between two halves of the confining potential, at $x=0$.
(\bf{b}) Time dependence of the ensemble average of the energy flux. Notice the scale difference between the values of $j_E$ in Fig. 4(a) and 4(b). (\bf{c}) The histogram of the distribution of the values of the advective component $\eta j_p$ of the energy flux between two halves of the confining potential.
(\bf{d}) Time dependence of the ensemble average of the advective component of the energy flux.
(\bf{e}) The histogram of the distribution of the values of the thermal component $j_{th}$ of the energy flux between two halves of the confining potential.
(\bf{f}) Time dependence of the ensemble average of the thermal component of the energy flux.}
\end{figure*}

Similarly we can consider 
\be
P_{1/2}(t)\equiv \int_{-\infty}^0 P (x,t)dx,
\ee
which determines the probability to find a particle at $x\leq 0$. Integrating Eq. (2) we get
\be
\left.\f{dP_{1/2}}{dt}=- j_p (t)\right |_{x=0}.
\ee
Figure 3 also shows the time dependence of ensemble average $\bar{P}_{1/2}$. Obviously, the probability does not accumulate in one half of the confining potential, so that there are numerous time intervals over which the time average probability flux 
is zero, $(\bar{P}_{1/2}(t_2)-\bar{P}_{1/2}(t_1))/(t_2-t_1)=0$, but the time average energy flux is not zero, 
$(\bar{E}_{1/2}(t_2)-\bar{E}_{1/2}(t_1))/(t_2-t_1)\neq 0$.
One can notice that there is small imbalance between the two halves of the potential well. The region where the perturbation is located becomes somewhat "deeper", so that on average $\bar{P}_{1/2}<0.5$, but the imbalance is small
\be
\f{\bar{P}_{1/2}-0.5}{0.5}\approx -0.02.
\ee

For the ensemble of solutions of Eq. (24) $\{\Psi_1(x,t)\ldots\Psi_N(x,t)\}$ we calculate the corresponding ensembles of fluxes 
$\{j_p^{(1)}(x,t)\ldots j_p^{(N)}(x,t)\}$ and $\{j_E^{(1)}(x,t)\ldots j_E^{(N)}(x,t)\}$ using Eqs. (3) and (9) at any given point as functions of time. 

A histogram in Fig. 4 (a) shows the normalized to unity probability distribution $dW(j_E)/dj_E$ for the energy flux across the centerline $x=0$.  
The natural units for the energy flux in one-dimensional system is $\hbar\omega^2$. 
This distribution is not a normal Gaussian because the tails are "fatter" than normal. It can be described as a superposition of normal and 
Laplace distribution 
\be 
dW(x)\sim \exp(-|x|).
\ee
Figure 4(b) shows the ensemble average of the energy flux across the centerline. The averaging procedure is the same as in (47)
\be
\bar{j}_E(t)=N^{-1}\left. \sum_{i=1}^{N}j_E^{(i)}(t) \right |_{x=0}.
\ee
As expected, one can see that there is a negative bias in this stochastic quantity corresponding to the net energy flow from the right-hand side of the potential well where the perturbation is localized to the free of perturbation left-hand side. 
Notice the difference in scale between Fig. 4(a) and 4(b). The ensemble average of the energy flux is relatively small
\be
\f{\bar{j}_E}{\hbar\omega^2}\sim \pm (0.02 - 0.04),
\ee
while the energy flux in an individual member of the enemble, as evident from Fig. 4(a), is substantially greater
\be
\f{j^{(i)}_E}{\hbar\omega^2}\sim \pm 0.5.
\ee
The large stochastic fluctuations of $j_E$ within the ensemble cancel each other out, resulting in relatively small ensemble average net energy flow. This is the reason the distribution in Fig. 4(a) looks almost symmetric. 

It is important to identify two different types of energy transfer. The center of mass of the particle, as determined by the probability flux, randomly moves between two halves of the confining potential and carries with it a certain amount of energy. The average of this component of the energy flux is obviously zero. The energy flux of a different nature, uncorrelated with the particle movement, we will call, for lack of a better term, the thermal component of the energy flux.
For each member of the ensemble the energy flux can be presented as a superposition of two stochastic components 
\be
j_E=\eta j_p +j_{th},
\ee
where $\eta j_p$ is the advective component of the energy flux. In a steady state ensemble, $\eta$ would be a constant - the average energy of advective flow. In our ensemble with average energy rising, $\eta$ slowly increases with time. The second component $ j_{th}$ is the "thermal flux". We define the thermal flux as a part of the energy flux $j_E$ that is statistically orthogonal to the probability flux, namely
\be
\langle\langle j_pj_{th}\rangle\rangle =0.
\ee
Here we use double brackets to indicate a stistical averaging, which may include both time and ensemble averaging, and differentiate it from the quantum average as in Eq. (5). The condition of orthogonality is 
\be
\eta=\f{ \langle\langle j_pj_{E}\rangle\rangle }{\langle\langle j_p^2\rangle\rangle }.
\ee
Because of the orthogonality only one of the components, either the advection or the thermal component of the total energy flux, can have a non-zero average. In our case the average probability flux is zero, so that the net energy flux is part of the thermal component $j_{th}$. The advective component describes the energy flow associated with random sloshing of the particle around the potential well and is not associated with the heating of the system. 
It should also be noted that the condition of orthogonality, Eq. (59), is equivalent to the condition of minimization of the difference $j_E-\eta j_p$, namely
\be
\f{\p}{\p\eta} \langle\langle (j_E-\eta j_p)^2\rangle\rangle =0.
\ee

In Fig. 4(c) the distribution of the advective component is shown. This is a purely symmetric distribution. The value of the average energy $\eta $ carried by the probability flux increses with time as the total energy increases. Figure 4(d) shows the time dependence of the ensemble average advective component of the energy flux. 

Figure 4(e) shows the distribution of the thermal component. It reminds more closely an asymmetric Laplace distribution 
\be
\f{dW(j_{th})}{dj_{th}}\sim \exp (-|j_{th}|/j_{\pm})
\ee
with the the constant $j_{+}<j_{-}$. Even in the thermal component of the energy flux the fluctuations are substantially greater than the net average, so that $|j_{+}-j_{-}|\ll (j_{+}+j_{-})/2 $. In Fig. 4(f) the ensemble average of the thermal component (obtained similar to (55)) is shown. Here the fluctuations cancel each other out and the negative bias is very clearly pronounced.

In our model the amplitudes of fluctuations of the energy flux are substantially greater than its net average. Segregation of the thermal flux component from the total energy flux allows us to better resolve a small effect (non-zero net energy flux) on the background of the large fluctuations. 
\subsection{\label{sec:level1}Quantum Advection Modes as Energy Carriers\protect}

One of the main goals of this study is to determine how different QAM contribute to the overall energy flux.
As one can expect, some of the modes will carry probability (charge and/or mass) and energy in the direction of the average energy flux, while the other modes will provide the backflow and carry probability in the opposite direction. Thus, all these advection modes can be identified as the Landauer channels in the quantum system with discrete energy spectrum. 

As an example, let us consider conditions when the average, over a certain time interval, probability flux across a given point
is zero, but the average energy flux over the same time interval is non-zero. Furthermore, let us consider, for the sake of argument, that there are only two active QAM, $q_{\alpha}$ and $q_{\beta}$ and $\epsilon_{\alpha} <\epsilon_{\beta} $. Then, obviously, the fact that the average $j_p= q_{\alpha}+q_{\beta}=0$ requires that the direction of the low energy mode, $q_{\alpha}$, anticorrelates with the direction of the energy flux $j_E=\epsilon_{\alpha}q_{\alpha}+ \epsilon_{\beta}q_{\beta}\neq 0$.
The lower energy mode provides the backflow of the probability. The direction of the mode $q_{\beta}$ which carries greater energy per unit of the probability flux must correlate with the direction of the energy flux. When there are several activated modes, the correlation between the average direction of the energy flux and each individual mode becomes less than $100\%$, but it still must be present. 

\begin{figure}
\begin{tabular}{cc}
\includegraphics [width=.9\columnwidth]{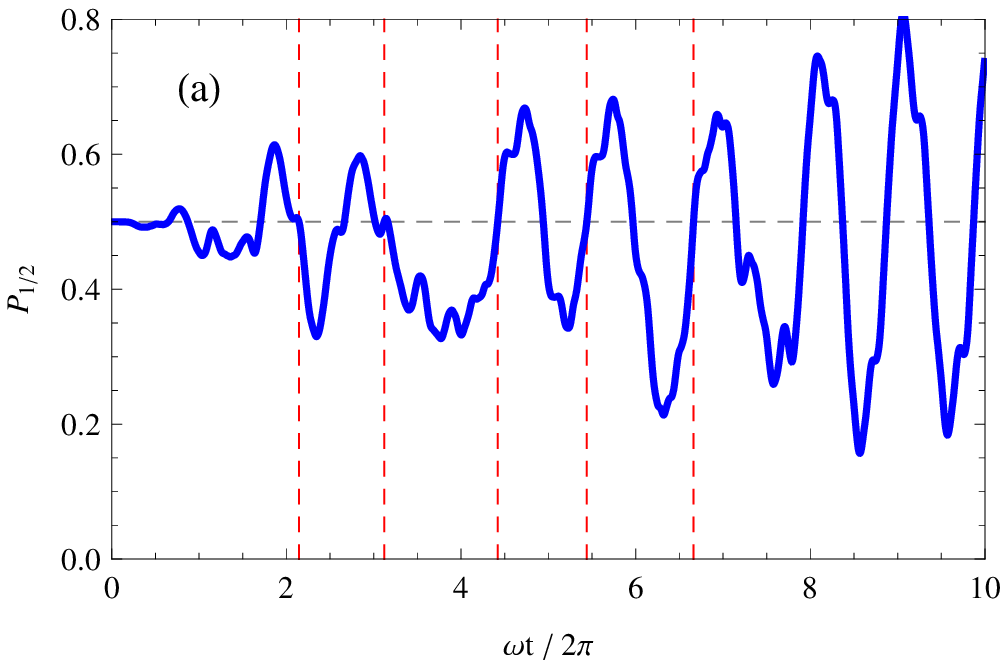} \\
\includegraphics [width=.9\columnwidth]{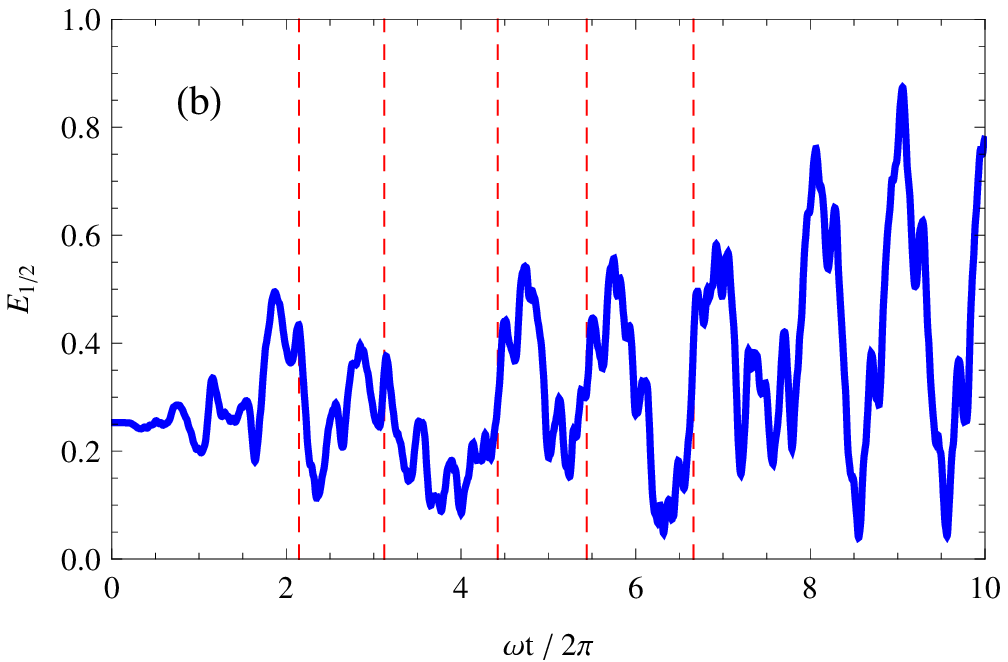}
\end{tabular}
\caption{\label{fig:} (\bf{a}) Time dependence of the half-bucket probability for an individual member of the ensemble, Eq. (51). 
(\bf{b}) Time dependence of the half-bucket energy, Eq. (49). Four consecutive averaging time intervals determined by the condition (63) are shown by the dashed lines. These intervals are different for every member of the ensemble.}
\end{figure}
\begin{figure}
\includegraphics{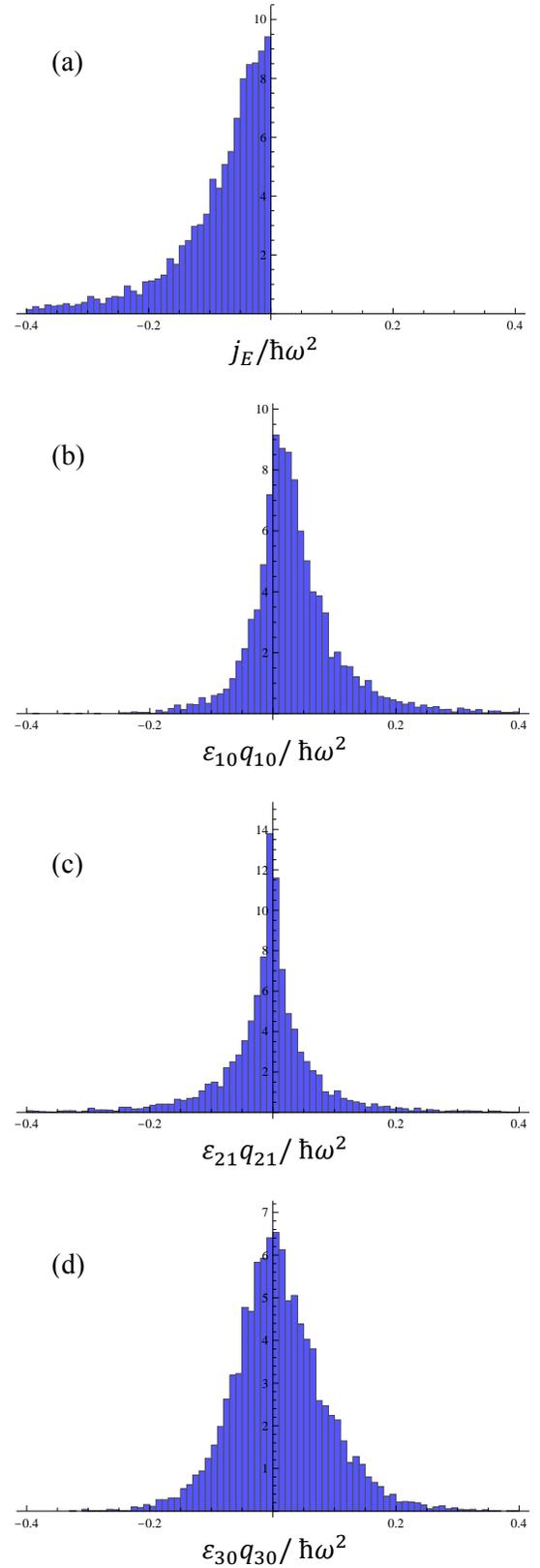}
\caption{\label{fig:} (\bf{a}) The distribution of the values of the energy flux $ \tilde{j}_{E}^i$ in the subensemble defined by a condition $ \tilde{j}_{E}^i\leq 0$. 
(\bf{b}) The distribution of the values of the lowest QAM $\tilde{q}_{10}^i$ in the same subensemble. 
(\bf{c}) The distribution of the values of the QAM $\tilde{q}_{21}^i$ in the same subensemble. 
(\bf{d}) The distribution of the values of the QAM $\tilde{q}_{30}^i$ in the same subensemble. }
\end{figure}
\begin{figure*}
\includegraphics{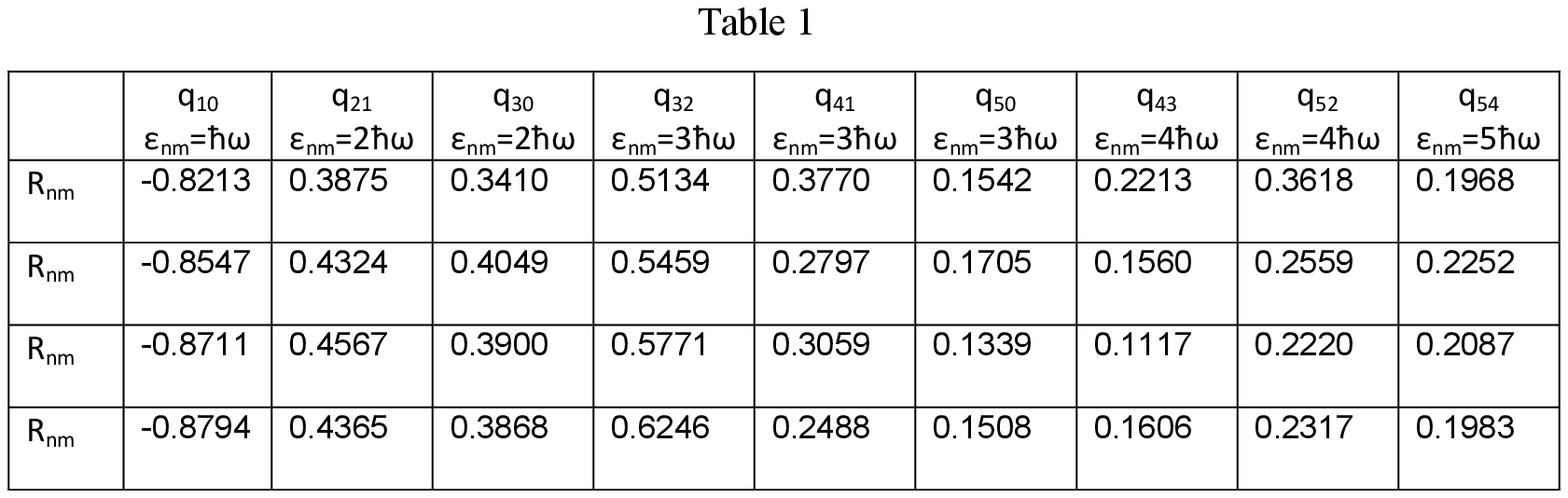}
\caption{\label{Table:}  
\bf{Table 1:} The values of the correlators $R_{nm}$, Eq. (68), for the first nine modes $\tilde{q}_{nm}$. The top row corresponds to the earliest time intervals. The other rows show the values of the correlators at progressively more distant times.}
\end{figure*}

Figure 5(a,b) shows a typical time variation of the half-bucket probability and energy, Eqs. (49) and (51), of an individual member of the ensemble. There are several instances when 
\be
P_{1/2}(t_k)=P_{1/2}(t_{k+1}),
\ee
so that the average, over the time interval $t_{k+1}-t_k$, probability flux is zero. However, the average energy flux over the same interval is, in most cases, nonzero
\bea
\tilde{j}_E^i(t_k,t_{k+1})\equiv \f{1}{t_{k+1}-t_k}\left. \int_{t_k}^{t_{k+1}}dtj_E^i(t)\right |_{x=0} =\\ \nonumber
=\f{E_{1/2}^i(t_{k+1})- E_{1/2}^i(t_{k})}{t_{k+1}-t_k}.
\eea
Here the superscript $i=1\ldots N$ indicates a member of the ensemble characterized by a wave function $\Psi_i(x,t)$.
The time intervals $t_k,t_{k+1}$ are different for different members of the ensemble. We have used several consecutive intervals progressively farther from the starting moment. 
The duration $t_{k+1}-t_k$ of the averaging intervals is of the order of one classical period, $\omega (t_{k+1}-t_k)/2\pi\sim 1 $.
For each member of the ensemble we determine the time intervals satisfying condition (63) and calculate the time average energy flux
given by Eq. (64) as shown in Fig. 5.

The average value of different QAM is calculated for each member of the ensemble by calculating the amplitudes $c_n(t)$, Eq. (11), and averaging the contribution of each mode 
\be 
q_{nm}^i(t) =\f{\hbar}{m}\sum_{n>m} \Im \left(c_nc_m^{\ast} e^{-i\omega_{nm} t}\right )^i\left.g_{nm}(x)\right |_{x=0}
\ee
over the the same time interval as in (64),
\be
\tilde{q}_{nm}^i\equiv 
\f{1}{t_{k+1}-t_k}\int_{t_k}^{t_{k+1}}q_{nm}^i(t)dt.
\ee

At $x=0$ only {\it odd-even} modes provide nonzero contribution because
\be
g_{nm}(x) =\psi_n^{\prime}(x)\psi_m(x) - \psi_n(x)\psi_m^{\prime}(x)
\ee
vanishes at $x=0$ if the indices $n,m$ are both odd or even. Only the modes with positive spatial parity contribute to the flow of matter and energy across the symmetry point.

A quantitative measure of the overall correlations between the energy flux (64) and a given mode (66) is determined by the correlators
\be
R_{nm}\equiv \f{\sum_{i=1}^N\tilde{q}_{nm}^i \tilde{j}_E^i}{(\langle \tilde {q}_{nm}^2 \rangle \langle \tilde{j}_{E}^2\rangle )^{1/2}},
\ee
where 
\be
\langle \tilde {q}_{nm}^2 \rangle =\sum_{i=1}^N( \tilde {q}_{nm}^i)^2;\; 
\langle \tilde {j}_{E}^2 \rangle =\sum_{i=1}^N( \tilde {j}_{E}^i)^2.
\ee

Figure 6 provides a qualitative view of the correlations between the energy flux and different modes. In Fig. 6(a) we show a histogram of a subensemble of the values of  $ \tilde {j}_{E}^i$ (Eq. (64)). This subensemble includes over half of the total number of members of the full ensemble for which $ \tilde {j}_{E}^i\leq 0$. The rest of the ensemble has $ \tilde {j}_{E}^i>0$. Figure 6(b) shows the distribution of the values of the lowest QAM $\tilde{q}_{10}^i$ for the same subensemble. This mode strongly anticorrelates with the direction of the energy flow and {\it on average} is responsible for the backflow of substance (probability flux) and energy in the direction opposite to the direction of the average energy flow.  We see exactly the same anticorrelation between $ \tilde {j}_{E}^i$ and $\tilde{q}_{10}^i$ in the rest of the ensemble for which 
$ \tilde {j}_{E}^i > 0$. 

Figure 6(c) shows the distribution of the values of $\tilde{q}_{21}^i$, one of the two degenerate modes which carry two quanta of energy per unit of the probability flux for the same subensemble. This mode clearly correlates with the energy flux and on average carries probability and energy in the same direction as the total energy flow. Figure 6(d) shows the distribution of the second degenerate mode $\tilde{q}_{30}^i$. For this mode the distribution appears more symmetric, so that its contribution to the net energy flow is somewhat smaller than that of $\tilde{q}_{21}^i$. Notice, that
for $\tilde{q}_{10}^i$ and $\tilde{q}_{21}^i$ the frequency $\omega_{nm}=\omega$, see Eq. (19), while for $\tilde{q}_{30}^i$ 
$\omega_{30}=3\omega$.

Table 1 shows the values of the correlators $R_{nm}$ (Eq. (68)) for nine lowest modes which account for about $90\%$ of the total
contribution to the values of $ \tilde {j}_{E}^i$. The averaging takes place over the whole ensemble. 
For each member of the ensemble
\bea
\tilde {j}_{p}^i=\sum_{n>m} \tilde{q}_{nm}^i=0;\\ \nonumber
\tilde {j}_{E}^i=\sum_{n>m} \epsilon_{nm}\tilde{q}_{nm}^i\neq 0.
\eea
The rows in the table correspond to successive time intervals like the ones shown in Fig. 5. The top row corresponds to the earliest time intervals and the other rows give the values of the correlators at progressively later time.

The values of the correlators confirm the qualitative picture seen in Fig. 6. The strongest correlation is exhibited by the lowest energy mode
$\tilde{q}_{10}^i$, which appears to be single-handedly carries the backflow of the probability flux equal to that carried by all other modes  in the opposite direction. This is the only mode that anticorrelates with the direction of the net energy flow in the large majority of the ensemble members. All other higher energy modes exhibit positive correlations with the direction of the energy flow. We expected that the lowest mode will be responsible for the backflow, but it is rather surprising that all other modes exhibit positive correlations within the statistical margins of error. The time progression does not change these correlators in any significant way. 

It is possible, that the choice of the ground state as the initial condition, which makes the correlator $c_0c_1^{*}$ greater than all other correlators, makes it inevitable that $q_{10}$ is the only mode responsible for the backflow. All other modes need to combine their contributions constructively in order to cancel that of $q_{10}$ to the total probability flux and prevail in determining the direction of the energy flux because they carry greater energy per unit of the probability flux. 

\section{\label{sec:level1}Summary and Speculations\protect}
The main purpose of this paper is to clarify the notion of Landauer channels in a system 
with discrete energy spectrum, whose eigenfunctions cannot be reduced to plane waves. 
We also wanted to address a situation in which there is no net probability (charge and/or mass) flux, but there is a net energy flux. 
In our model we produce the persistent energy flux by pumping energy into the system 
with the help of a localized and asymmetrically spaced power source. 

In the region free of the time-dependent perturbation the energy and probability flux can be 
defined as a superposition of quantum advection modes, each associated with an off-diagonal element of the density matrix.
Our results show that these modes play the role of Landauer channels, but in a way very different from their conventional interpretation.
First of all, each mode does not simply carry probability and energy in a certain direction. The direction and amplitude of the modes fluctuate, and only statistically correlate with the net energy flow. The lowest energy mode provides on average the backflow of the probability and energy. The higher energy modes have positive correlations with the energy flux. Thus, the total probability flux can be zero because the contribution of the backflow mode cancels out that of the higher energy modes. However, since the higher modes carry greater amount of energy per unit of the probability flux, they determine the value and the direction of the net energy flow.

Usually, the energy flow unaccompanied by the mass flow is defined as heat transfer. The results shown above indicate that a constant 
heat flux exists due to persistent coherence between the amplitudes of the different energy levels. As long as the external conditions that induce the persistent energy flow are maintained, the ensemble average values of the correlators $ c_nc_m^{\ast} e^{-i\omega_{nm} t}$ are not zero. 

This observation might have implications beyond the heat transfer topic. Decoherence of qubits, the fundamental elements of the hypothetical quantum computers, is a major obstacle on a way to their construction. 
One might suggest that a qubit placed in contact with the nonequilibrium environment, such that it induces a persistent energy flow through the qubit, will be protected to some extend from complete decoherence.
One can see the illustration of this point in Table 1. The mode  $ q_{10}\sim \Im \left(c_1c_0^{\ast} e^{-i\omega_{nm} t}\right )$ has over $80\% $ correlation with the net heat flux. Thus, strong persistent coherence between the ground and the first excited state exists for as long as we maintain the energy flow through this quantum system. We should mention again that trapped ions, one of the early candidates for the role of qubits, are well described by a model of a particle confined to a harmonic potential well and exposed to stochastic perturbation due to electromagnetic noise \cite{Turchette,Turchette2, Leib}. 
Of course, at this point this is only a speculation and more work needs to be done to determine whether this hypothesis has any merit.

\section{\label{sec:level1}Acknowledgment\protect}
This work was supported by the Air Force Office of Scientific Research. 
We thank Andrew Sommers, George Panasyuk, and Tim Haugan for many helpful
discussions.

\section{\label{sec:level1}Appendix: Derivation of Energy Continuity Equation\protect}
There are many ways one can crumble this cookie. Here we will show that Eqs. (7) and (9) are correct.
The energy density defined by Eq. (4) and the Hamiltonian (1) is given by
\be
\epsilon=-\f{\hbar^2}{2m}\Re\left (\Psi^{\ast}\nabla^2\Psi\right )+
V|\Psi |^2\equiv \epsilon_1 + \epsilon_2.\nonumber
\ee 
The rate of change
\be
\f{\p\epsilon }{\p t}\equiv \dot{\epsilon}_1+\dot{\epsilon}_2,\nonumber
\ee
where
\be
\dot{\epsilon}_1=-\f{\hbar^2}{2m}\Re\left (\dot{\Psi}^{\ast}\nabla^2\Psi 
+\Psi^{\ast}\nabla^2\dot{\Psi}\right ),\nonumber
\ee
and
\be
\dot{\epsilon}_2=\dot{V}|\Psi |^2 +V\f{\p |\Psi |^2}{\p t}.\nonumber
\ee
We can rewrite $\dot{\epsilon}_1$ in the form
\be
\dot{\epsilon}_1=-\f{\hbar^2}{2m}\Re\left (\Psi^{\ast}\nabla^2\dot{\Psi}-\dot{\Psi}^{\ast}\nabla^2\Psi 
+2\dot{\Psi}^{\ast}\nabla^2\Psi \right ).\nonumber
\ee
Then, it is easy to see that
\bea
-\f{\hbar^2}{2m}\Re\left (\Psi^{\ast}\nabla^2\dot{\Psi}-\dot{\Psi}^{\ast}\nabla^2\Psi \right )= \nonumber \\
-\f{\hbar^2}{2m}\Re\left \{\nabla\cdot ( \Psi^{\ast }\nabla\dot{\Psi }-\dot{\Psi } \nabla \Psi^{\ast})\right \}\equiv 
-\nabla\cdot \vec{j}_E.\nonumber
\eea
Now we only need to show that
\be
V\f{\p |\Psi |^2}{\p t}- \f{\hbar^2}{m}\Re \{\dot{\Psi}^{\ast}\nabla^2\Psi \}\equiv 0.\nonumber 
\ee
This condition is equivalent to the probability conservation condition (2). Indeed, by virtue of Eq. (1)
\bea
-\f{\hbar^2}{m}\Re \{\dot{\Psi}^{\ast}\nabla^2\Psi \}=\nonumber
\f{\hbar}{m}\Im(\hat{H}\Psi^{\ast}\nabla^2 \Psi ) = \\ \nonumber
\f{\hbar}{m}V\Im (\Psi^{\ast}\nabla^2\Psi ) = 
\f{\hbar}{m}V\Im\{ \nabla\cdot (\Psi^{\ast}\nabla\Psi )\}\equiv V\nabla\cdot \vec{j}_p. \nonumber
\eea

If we take the wave function $\Psi$ in the form
\be
\Psi (\vec{r},t)=Re^{iS},\nonumber 
\ee
so that 
\be
\vec{j}_p=\f{\hbar}{m}R^2\nabla S, \nonumber 
\ee
we get 
\be
\vec{j}_E= \f{\hbar^2}{2m}\{R\nabla\dot{R}-\dot{R}\nabla R\}-\hbar\dot{S}\vec{j}_p. \nonumber 
\ee

\bibliography{apssamp}

%
%
\end{document}